\documentclass[twocolumn,superscriptaddress,showpacs,
showkeys,preprintnumbers,amsmath,amssymb]{revtex4}
\usepackage{graphicx}
\usepackage{dcolumn}
\usepackage{bm}
\begin{document}
\title{Noise induced currents and reliability of transport in frictional
ratchets}
\author{Raishma Krishnan}
\affiliation{Institute of Physics, Bhubaneswar-751005, India}
\email{raishma@iopb.res.in, jayan@iopb.res.in}
\author{Debasis Dan}
\affiliation{Department of Physics, Indiana University, Bloomington 
47405, U.S.A}
\email{ddan@indiana.edu}
\author {A. M. Jayannavar}
\affiliation{Institute of Physics, Bhubaneswar-751005, India}
\begin{abstract}
We study the coherence of transport of an overdamped 
Brownian particle in frictional ratchet system in the presence 
of external Gaussian white noise fluctuations. The analytical expressions 
for the particle velocity and diffusion coefficient are derived 
for this system and the reliability or coherence of transport is analysed 
by means of their ratio in terms of a dimensionless P$\acute{e}$clet number. 
We show that the coherence in the transport can be enhanced or degraded 
depending sensitively on the frictional profile with respect to the underlying
potential.
\end{abstract}
\pacs{05.40.Jc, 05.40.Ca, 02.50.Ey}
 \keywords{Ratchets, Brownian motors, Noise, Transport coherence}
\maketitle

\section{Introduction}

The study of the interplay of noise and nonlinear dynamics in systems 
under nonequilibrium conditions has generated 
wide interdisciplinary interests in the last two decades. 
Noise or fluctuations, which are normally considered to be an hinderance, 
are found to play an active constructive 
role in nonequilibrium systems. Infact, this constructive role of noise (as
opposed to the conventional wisdom of its destructive or its 
disorganising role) has become a new paradigm in the study 
of complex systems. In the so called ratchet systems the 
presence of spatial/temporal anisotropy in potential together 
with nonequilibrium perturbations enable the extraction 
of useful work from random fluctuations without the violation 
of the second law of thermodynamics~\cite{reiman,1amj}. In such systems it 
is possible to induce directed motion from nonequilibrium 
fluctuations in the absence of bias.  Much of the studies 
in different classes of ratchet models deal with the  
nature of currents and their reversals~\cite{reiman}, stochastic energetics 
(thermodynamic efficiency)~\cite{parrondo,pre-rk} etc. However, transport 
of Brownian particle is always accompanied by 
a diffusive spread and this spread is intimately related 
to the question of reliability or quality of transport. The 
diffusive spread infact detriments the quality of transport. 
There exists very few studies which address the question 
of diffusion accompanying transport in ratchet 
systems~\cite{low,sch,ijp}. In our present 
work we address this aspect of transport and study the 
coherence in transport in a frictional ratchet 
in the presence of an external parametric Gaussian white noise 
fluctuation. This is infact studied in terms of a dimensionless 
quantity called the P$\acute{e}$clet
number $(Pe)$ which is the ratio of velocity to the 
diffusion constant. Higher the $Pe$, lesser is the diffusive 
spread and higher is the transport coherence. Infact, 
subcellular transport in biological systems amidst 
a noisy environment are modeled based on the principle 
of ratchet mechanism and the experimental studies on these 
molecular motors show them to have highly efficient and reliable transport 
with P$\acute{e}$clet number ranging from 
2 to 6~\cite{high}. A value of $Pe$ greater than $2$  
corresponds to coherent transport~\cite{low}. The P$\acute{e}$clet numbers for 
some of the models like flashing and rocking ratchets were found to be 
$ \sim 0.2$ and $ \sim 0.6$~\cite{low} respectively implying a less reliable 
transport. Another study on symmetric periodic potentials 
along with spatially modulated white noise showed a coherent transport with 
P$\acute{e}$clet number less than $3$. In the same study a 
special kind of strongly asymmetric potential is found to 
increase $Pe$ to $20$ in some range of physical parameters~\cite{sch}. 

There exists many physical systems like flashing 
ratchets~\cite{ajdari}, rocking ratchets~\cite{magnasco}, 
time asymmetric ratchets etc., where different aspects of noise 
induced transport has been widely studied~\cite{reiman,1amj}. In 
the above models, to generate unidirectional current the nonequilibrium 
fluctuations need to be correlated in time. 
There exists a possibility to get unidirectional current 
even in  presence of symmetric ratchet potentials provided it is 
driven by a time correlated asymmetric force~\cite{rock-prl}. 
In our present work we consider yet another class of ratchets, namely the
frictional ratchets, where the friction 
coefficient and subsequently the diffusion coefficient 
varies in space~\cite{amj,pareek}. In these frictional 
ratchets  it is possible to get unidirectional currents even 
in a symmetric underlying potential but in presence of an 
external noise which need not be correlated in time unlike 
in earlier models. 

Space dependent diffusion coefficient, $D(x)$, felt by the Brownian particle 
could arise either due to space dependent temperature or space 
dependent friction coefficient. In
the frictional ratchet system which we consider in the present work the
unidirectional current arises due to a combination of both space dependent 
friction coefficient and external parametric white noise. 
The temperature of the bath (or the environment) of the 
Brownian particle is characterised by a constant temperature $T$.  
In the presence of external parametric noise 
 the overdamped Brownian particle on an average absorbs energy 
from the external noise source. The strength of the 
absorbed energy depends on the local
frictional coefficient. Hence the problem 
 of particle motion in an inhomogeneous medium in presence of 
 an external noise becomes equivalent to the problem in a space 
 dependent temperature~\cite{amj,pareek,buttiker}. Such systems 
are known to generate unidirectional currents. 
 This  follows as a corollary to  Landauer's blow torch theorem 
that the notion of stability changes 
 dramatically in the presence of temperature inhomogeneities~\cite{land}. 
 In such cases the notion of local stability, valid in equilibrium 
 systems, does not hold. 

Frictional inhomogeneities are common in superlattice structures, 
 semiconductors or motion in porous media. Particles moving close to a surface
experience space dependent friction~\cite{surface}. It is believed that 
molecular motor protiens moving close along the periodic 
structures of microtubules  experience a space dependent
friction~\cite{luch}. Frictional inhomogeneity changes 
the dynamics of the particle nontrivially as 
 compared to the homogeneous case. This in turn has been shown to 
 give rise to many counter intuitive phenomena like 
noise induced stability, stochastic resonance, enhancement in efficiency 
etc.,  in driven non-equilibrium systems~\cite{luch,thesis}.


In our present work we show that system inhomogeneities may help in 
enhancing/degrading the coherence in the transport depending sensitively on 
the physical parameters.  We emphasize mainly the case where the underlying
potential is a simple sinusoidal symmetric potential. The role of spatial
asymmetry in potential is also discussed. The external noise 
is found to  play a constructive role in enhancing 
the coherence in transport. As opposed to this, 
temperature (internal fluctuations) degrades the coherence in transport.  

\section{Model:}
 
We start with the Kramer's equation of motion for a Brownian 
particle of unit mass in contact with a heat bath in a medium with 
spatially varying friction coefficient $\eta(q)$ at temperature $T$. In 
addition an external parametric Gaussian white 
noise fluctuation $\xi(t)$ is also
included. The equation of motion is given by 
\begin{equation}
 {\ddot{q}} =-\eta(q){\dot{q}}- { {{V^\prime(q)}}} 
+ {\sqrt { {k_BT}{\eta(q)}}}f(t)+\xi(t) \label{leqn}
 \end{equation}
where $V(q)$ is the potential seen by the Brownian particle and $f(t)$ is an
internal Gausian white noise fluctuation arising from the bath having the
property that $<f(t)> = 0$, and $ <f(t)f(t^\prime)> = 2 \delta(t-t^\prime)$ 
 where $<...>$ 
 denotes the ensemble average and $q$ the coordinate of the particle. Also,
$<\xi(t)>=0$ and $<\xi(t) \xi(t^\prime)> = 2 \Gamma \delta(t-t^\prime)$,
 where $\Gamma$ is the strength of the external white noise $\xi(t)$. The above
equation, Eq.~\ref{leqn}, has been derived earlier from the microscopic
consideration of system bath coupling~\cite{amj,pareek}.   

On time scales larger than the inverse friction
coefficient, $\eta^{-1}$, one can in most practical 
cases consider the overdamped limit of the Langevin
equation. This in turn correspond to the adiabatic elimination of 
the fast variable, velocity, from the equation of motion by putting
$\dot{p}=\ddot{q} = 0$ for a homogeneous system. In contrast, for the case of
inhomogeneous system the above method of elimination does not work and 
Sancho et al.~\cite{sancho} has given a proper prescription 
for the elimination of fast variables.  The corresponding overdamped 
Langevin equation for the Brownian particle in a space dependent 
frictional medium is given by

\begin{equation}
 \dot{q} = {- \frac{V^\prime(q)}{\eta(q)}} - \frac {k_BT {\eta^\prime}
 (q)}{ 2{[\eta(q)]}^2} + {\sqrt{\frac{k_BT}{\eta(q)}}f(t)} + 
\frac{\xi(t)}{\eta(q)}. \label{langevin}
 \end{equation}

 Using van Kampen Lemma~\cite{vankampenlemma} and Novikov's 
theorem~\cite{novikov} we get the corresponding Fokker-Planck or 
Smoluchowski equation for the probability density $P(q,t)$ of a particle being
at $q$ at a time $t$ as \cite{risken}
\begin{equation}
 \frac {\partial P}{\partial t}=  \frac {\partial}{\partial q} 
\left[{\left\{\frac{V^\prime(q)}{\eta(q)}-\frac{\Gamma \eta^\prime(q)}{\eta^3(q)}\right\}P  + \left\{\frac{k_BT}
 {\eta(q)}+ \frac{\Gamma}{\eta^2(q)}\right\}\frac{\partial P}{\partial q}}\right]\,.
 \end{equation}

 For periodic functions $V(q)$ and $\eta(q)$ with periodicity $L$ 
a finite probability current is
obtained and one readily gets the analytical expression for 
the particle velocity  as ~\cite{risken}
 \begin{equation}
 v= L\frac{(1-\exp\,({-\delta}))}{\int_0^{2\pi} dy \exp\,[-\psi(y)] 
 \int_{y}^{y+2\pi}dx \frac{\exp\,\,[{\psi(x)}]}{A(x)}}.\label{current}
 \end{equation}
Here  $\psi(q)$ is the dimensionless generalized effective potential given by  
 \begin{equation}
 \psi(q)=\int^{q} dx \frac{V^\prime(x)\eta^2(x)-\Gamma\eta^\prime(x)}{\eta(x)
[k_BT\eta(x)+\Gamma ]}\label{eff-pot}
 \end{equation}
 and  $A(q)$ is the effective space dependent diffusion coefficient given by
 \begin{equation}
 A(q)=\frac{k_BT \eta(q)+\Gamma}{\eta^2(q)}.\label{aq}
 \end{equation}
 Also, 
\begin{equation}
\delta = \psi(q)-\psi(q+2\pi) \label{delta}
\end{equation}
determines the effective slope of the generalized  potential $\psi(q)$. 
Thus the sign of $\delta$ gives the 
 direction of current in Eq.~\ref{current}. 

 In our present work we have taken the potential $V(q)=V_0 \sin(q)$
and $\eta(q)=\eta_0[1-\lambda 
\sin(q-\phi)]$, $0\,<\,\lambda\,<\, 1$. The phase lag $\phi$ between 
$V(q)$ and $\eta(q)$ brings in the intrinsic asymmetry in 
the dynamics of the system.
 
For the case with $T=0$ we get a simple analytical expression for the
effective potential as 
\begin{eqnarray}
\psi(q)&=&\frac{V_0\eta_0}{\Gamma}\Big(\frac{\lambda x \sin(\phi)}{2}+\frac{\lambda}{4}\big[\cos(2x)\cos(\phi)\\
 & & +\sin(2x)\sin(\phi)\big]+\sin(x)\Big)-\ln[\eta(x)] \label{zeroTeffpot}
\end{eqnarray}
with $\delta$ given by
\begin{eqnarray}
\delta=-\frac{V_0 \eta_0 \pi \lambda \sin(\phi)}{\Gamma}.\label{deltazeroT}
\end{eqnarray}

The first term in the right hand side of Eq.~\ref{zeroTeffpot} represents the
tilt in the effective potential. This tilt identically vanishes when $\phi=0$ or
$\pi$. Hence it is expected that the unidirectional current does not arise for
the case when the phase lag is $0$ or $\pi$. From Eq.~\ref{current} and
Eq.~\ref{deltazeroT} it is clear that for $0 < \phi < \pi$ 
current will be in the negative direction while for  
$\pi < \phi < 2\pi$ the current will be in the positive direction.
\begin{figure}[h]
  \centering
\begin{center}
\input{epsf}
\hskip15cm \epsfxsize=2.5in \epsfbox{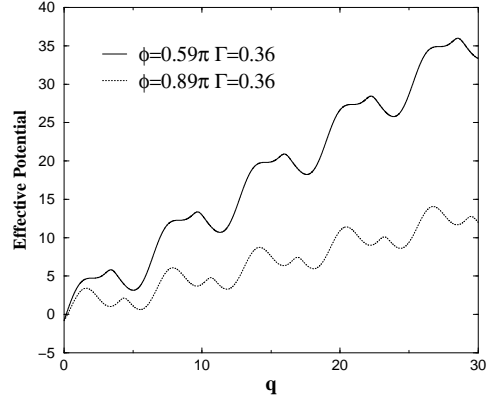}
   \caption{Effective potential $\psi(q)$ as a function of $q$ 
for $V_{0}=1$, $\Gamma=0.36$ and $\lambda=0.9$ at zero temperature 
for $\phi = 0.59\pi$ and $0.89\pi$.} \label{effpotential}
\end{center}
\end{figure}
Fig.~\ref{effpotential} shows the  plot of the effective potential 
as a function of coordinate $q$ for a fixed noise strength $\Gamma=0.36$ 
for two different values of  phase lag $\phi$ between the potential and 
frictional profile for the above zero temperature case. The effective
potential is scaled with respect to $V_0$. For the
parameters chosen in the figure the current flows in the negative 
direction as has been mentioned earlier.\\

 Following references ~\cite{rec,dan-giant}, one can obtain exact 
analytical expressions  for the diffusion coefficient $D$ as
 \begin{equation}
 D=\frac{\int_{q_0}^{q_0+L}\frac{ dx}{L}\,A(x)\, 
{[I_+(x)]}^2 I_-(x)}{\left[{\int_{q_0}
 ^{q_0+L}\frac{dx}{L}I_+(x)}\right]^3}\label{diffusion}
 \end{equation}
  where $I_+(x)$ and $I_-(x)$ are as given below
 \begin{eqnarray}
 I_+(x)&=& \frac{1}{A(x)}\,\,\exp\,[\psi(x)]\,\int_{x-L}^{x} dy 
 \,\,\exp\,[-\,\psi(y)] \, , \label{iplus} \\
 I_-(x)&=& \exp\,[- \psi(x)] \int_{x}^{x+L} dy\,\, 
 \frac{1}{A(y)}\,\exp\,[\psi(y)] \, .\label{iminus}
 \end{eqnarray}
 $L$ here represents the period of the potential ($=2 \pi$ in our case). 
The Brownian particle takes a time $\tau=L/v$ to traverse a distance $L$ with a
velocity $v$. The diffusive spread of the particle in the same time is given by $<(\Delta q)^2>=2D\tau$.  
The criterion to have a reliable transport is that the diffusive spread should
be less compared to the distance traversed, i.e.,  
 $<(\Delta q)^2>=2D\tau < L^2$. This in turn implies that $Pe=Lv/D >2$ 
 for coherent transport. 

 \section{Results and Discussions}

 The velocity ($v$), diffusion constant ($D$) and the 
 P$\acute{e}$clet number ($Pe$) are studied  as a function 
of different physical parameters. All the physical quantities are taken 
in dimensionless form. In particular, velocity and diffusion are 
normalized by $(V_0/\eta_0 L)$ and $(V_0/\eta_0)$ 
respectively. Throughout our work we have set $V_0$ and $\eta_0$ 
to be unity. Similarly, $\Gamma$ and $T$ are scaled with
respect to $V_0\eta_0$ and $V_0$ respectively. We have used the globally 
adaptive scheme based on Gauss-Kronrod rules for numerical
evaluations~\cite{gauskronrod}.\\ 

\begin{figure}[htp!]
 \begin{center}
\input{epsf}
\hskip15cm \epsfxsize=2.5in \epsfbox{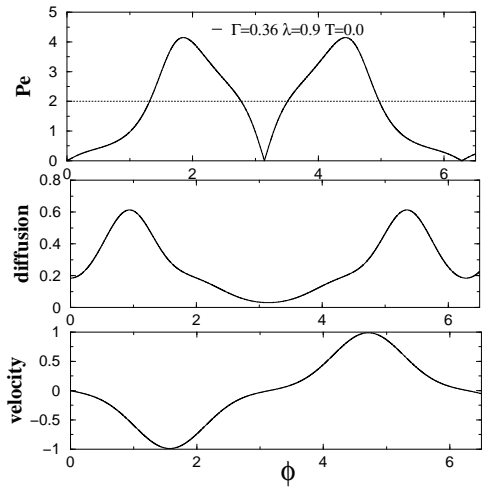}
\caption{Plot of v, D and Pe vs $\phi$ for $\Gamma=0.36$, 
$\lambda=0.9$ and $T=0.0$. } \label{m0t0vsphi}
 \end{center}
 \end{figure}
In Fig.~\ref{m0t0vsphi} we plot $v$, $D$ and  $Pe$ as a function of
phase difference $\phi$ at $T=0$ for a fixed noise strength of 
$\Gamma=0.36$. The values of the physical parameters are mentioned in 
the figure caption. All the physical quantities are periodic 
function of $\phi$ as expected and the velocity is zero 
at $\phi=0\,,\,\pi$ and $2\pi$. As can be seen 
from the plot of effective potential, Fig.~\ref{effpotential}, 
the direction of current is  negative for $\phi$ upto $\pi$ 
and then it becomes positive. The current is antisymmetric around the point
$\phi=\pi$, $(V(\phi+\pi)=-V(\phi))$. This is expected on general grounds
for the case of a simple sinusoidal symmetric potential. The absolute value of
current exhibits a maxima between $0$ to $\pi$ and $\pi$ to $2\pi$. The nature
of currents being positive or negative can be readily inferred from the slope of
the effective potential (for example see Fig.~\ref{effpotential}). The diffusion
coefficient is finite at all values of $\phi$ and exhibits minima wherever the
currents are zero $(\phi=0,\pi,2\pi)$. However, the magnitude of minima is more
at the point where $\phi=\pi$. As expected, the diffusion coefficient 
is periodic in $\phi$. However, it exhibits maxima at different values of 
$\phi$ than that for current. In the plot for $Pe$ as a function 
of $\phi$ we have drawn a dotted line as a guide for eye. 
In regions where $Pe > 2$  the transport is said to be coherent. It is evident 
from the figure that there exists wide range of $\phi$ in which
the transport is coherent $(Pe > 2)$ along with adjoining regions where the
transport is less coherent $(Pe < 2)$. These regions are however sensitively
dependent  on other physical parameters. In the present case of sinusoidal
potential P$\acute{e}$clet number as high as $4$ can be obtained by properly
tuning the parameters. 

\begin{figure}[htp!]
 \begin{center}
\input{epsf}
\hskip15cm \epsfxsize=2.5in \epsfbox{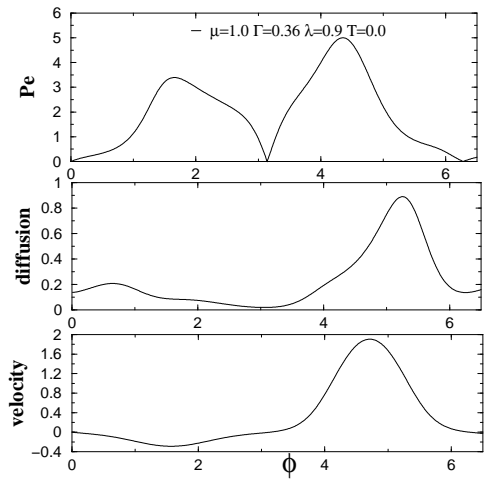}
 \caption{Plot of v, D and Pe vs $\phi$ for $\Gamma=0.36$, $\mu=1.0$, 
$\lambda=0.9$ and $T=0.0$.} \label{m1t0vsphi}
 \end{center}
 \end{figure}


To understand the role of spatial asymmetry in the potential 
we have included an asymmetry in the potential such that 
$V(q)=V_0[\sin(q) - \mu/4 \sin(2q)]$ where $\mu$ is the asymmetry 
parameter. Fig.~\ref{m1t0vsphi} shows the behaviour 
of $v$, $D$ and $Pe$ for this simple asymmetric case at zero temperature 
with the other physical parameters kept the same as in 
Fig.~\ref{m0t0vsphi}. The potential asymmetry parameter $\mu$ affects the
behaviour of velocity, diffusion and thereby $Pe$ dramatically. We 
have considered the case of maximal asymmetry $(\mu=1)$. We first notice that
the simple symmetry observed for the case of 
periodic potential no more holds true in the presence 
of spatial asymmetry in potential. The magnitude of 
velocity is nonzero for $\phi$ values $0\,,\,\pi$ and $2\pi$. The 
velocity and diffusion constant in the region between $0$ to $\pi$ 
are suppressed while in the region between $\pi$ and $2\pi$ 
are enhanced as compared with symmetric case. 
Consequently, the presence of asymmetry can enhance or
suppress the coherence in the transport. For the 
chosen parameters, P$\acute{e}$clet number as high 
as $5$ is obtained. Moreover, for a given $\phi$, 
$Pe$ either increases or decreases  monotonically 
with the asymmetry parameter $\mu$ $(0<\mu<1)$ which 
has been verified separately. In our further 
analysis we restrict to the case of symmetric potential alone.

\begin{figure}[hbp!]
 \begin{center}
\input{epsf}
\hskip15cm \epsfxsize=2.5in \epsfbox{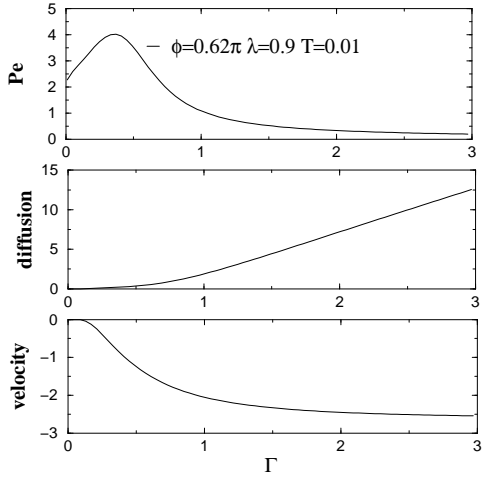}
 \caption{Plot of v, D and Pe vs $\Gamma$ for 
$\phi=0.62\pi$, $\lambda=0.9$ and $T=0.01$. } \label{m0phi62t01-G}
 \end{center}
 \end{figure}
In Fig.~\ref{m0phi62t01-G} we plot velocity, diffusion and $Pe$ 
as a function of external noise strength $\Gamma$ with $\phi=0.62 \pi$ and 
$\lambda=0.9$ at finite temperature $T=0.01$. It should be noted that velocity 
is negative in the entire range. The velocity is initially zero for $\Gamma$
equal to zero and then increases with $\Gamma$ and saturates to a constant
value at higher values of noise strength. On the contrary, 
the diffusion constant keeps increasing monotonically with $\Gamma$. 
It is also clear that the external noise
play a constructive role in optimizing the coherence in transport 
i.e., the $Pe$ exhibits a peak as a function of $\Gamma$. 

\begin{figure}[htp!]
 \begin{center}
\input{epsf}
\hskip15cm \epsfxsize=2.5in \epsfbox{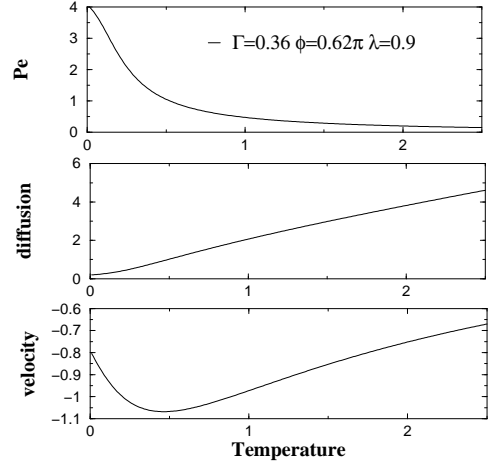}
 \caption{Plot of v, D and Pe vs $T$ for $\Gamma=0.36$, 
$\phi=0.62\pi$ and $\lambda=0.9$. } \label{m0phi62G36-T}
 \end{center}
 \end{figure}
\begin{figure}[htp!]
 \begin{center}
\input{epsf}
\hskip15cm \epsfxsize=2.5in \epsfbox{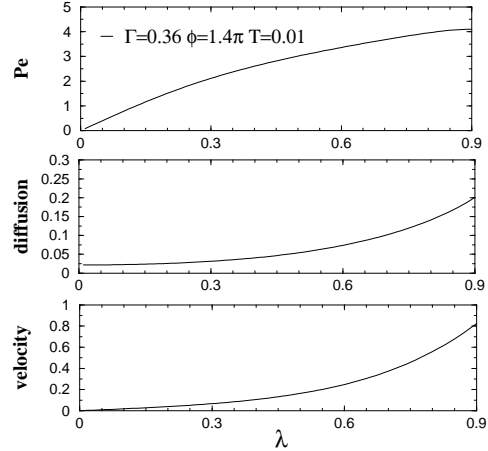}
 \caption{Plot of v, D and Pe vs $\lambda$ for $\Gamma=0.36$, 
$\phi=1.4\pi$ and $T=0.01$. }\label{m0phi14-lam}
 \end{center}
 \end{figure}

In Fig.~\ref{m0phi62G36-T} we plot velocity, diffusion and $Pe$ 
as a function of temperature $T$ with $\phi=0.62 \pi$, $\lambda=0.9$ 
and $\Gamma=0.36$. This value of noise strength corresponds to an optimal value
for $Pe$ in Fig.~\ref{m0phi62t01-G}. The noise induced current 
is negative. The current exhibits a peak and then decreases 
to zero with temperature. This is expected because 
higher temperatures overshadows the effect of potential 
and frictional inhomogenieties thereby 
suppressing the ratchet effect. In contrast, the diffusion constant increases
monotonically. As opposed to the case of Fig.~\ref{m0phi62t01-G} we observe that
internal fluctuations or temperature degrades the coherence in the transport
i.e., $Pe$ decreases with increase in temperature.\\

In Fig.~\ref{m0phi14-lam} we plot velocity, diffusion and $Pe$ 
as a function of  $\lambda$, the amplitude of oscillation of the friction
coefficient $(0<\lambda<1)$ for $\phi=1.4 \pi$, $\Gamma=0.36$ and $T=0.01$. 
All the physical quantities, namely, velocity, diffusion and $Pe$ increases
monotonically with $\lambda$. Thus the increase in $\lambda$ 
makes the transport more coherent.\\

\section{Conclusions}

We have studied the coherence or reliablity of transport of an overdamped
Brownian particle in a frictional ratchet system with an underlying sinusoidal
potential in the presence of external Gaussian white noise fluctuations. 
The frictional inhomogeneities along with external fluctuations lead to a noise
induced current or transport. The attained noise induced transport is always 
accompanied by a diffusive spread which inturn makes the transport to be less
reliable. We have shown that frictional inhomogeneities
with respect to the underlying potential can make the transport coherent or
incoherent. While the external noise $(\Gamma)$ optimizes 
the coherence in transport the internal noise $(T)$ degrades the coherence.  

In our present case, as mentioned in the beginning, the transport can be
associated with an effective potential and an effective space dependent 
diffusion constant. The effective potential exhibits a tilt as a function of
system parameters (Fig.~\ref{effpotential}). By looking at this effective
potential one can infer only the direction of current and not its magnitude. The
particle motion in this effective potential is determined by two time scales,
(i) escape from the potential minima over the barrier along the effective bias
followed by (ii) the relaxation into next minima. The coherent transport in
homogeneous medium is obtained when the relaxation time dominates the
transport~\cite{sch,dan-giant} as compared with the escape time. For details see
reference~\cite{sch,dan-giant}. We would like to emphasize that the dynamics of the particle in our present problem arises due to the complex 
interplay between the potential, internal fluctuations, 
frictional profile and external fluctuations.
This is amply reflected in the fact that both the effective potential and space
dependent diffusion constant change their nature as we change the system
parameters. Thus apriori analysis or prediction of the behaviour of the system 
in regard to transport coherence is a difficult task. It may not be surprising
that by choosing appropriate asymmetric potential and frictional profile along
with other parameters one may obtain much higher transport coherence as observed
in~\cite{sch}.   

 \end{document}